\documentclass[pre,twocolumn,showpacs,dvipdf,amsmath,amssymb,floatfix]{revtex4}
\usepackage{bm}
\usepackage{graphicx}

\begin{document}

\newcommand{\rbar}{\bar{R}}
\newcommand{\rbarmin}{\bar{R}_{\mathrm{min}}}
\newcommand{\rmin}{R_{\mathrm{min}}}
\newcommand{\rmindot}{{{\dot R}_{\mathrm{min}}}}
\newcommand{\zmin}{z_{\mathrm{min}}}
\newcommand{\sigtwo}{S_2}
\newcommand{\sigthree}{S_3}
\newcommand{\sign}{S_n}


\title{Perturbed breakup of gas bubbles in water: 
Memory, gas flow, and coalescence}

\author{Nathan~C.~Keim}
\email{nkeim@uchicago.edu}
\affiliation{James Franck Institute and Department of Physics, University of Chicago,
929 E.\ 57th St., Chicago, IL 60637, USA}
\altaffiliation[Present address: ]{Department of Mechanical Engineering and Applied Mechanics, University of Pennsylvania, 220 S.\ 33rd St., Philadelphia, PA 19104}

\date{\today}

\begin{abstract}

The pinch-off of an air bubble from an underwater nozzle ends in a singularity with a remarkable sensitivity to a variety of perturbations. I report on experiments that break both the axial (\textit{i.e.}, vertical) and azimuthal symmetry of the singularity formation. The density of the inner gas influences the axial asymmetry of the neck near pinch-off. For denser gases, flow through the neck late in collapse changes the pinch-off dynamics. Gas density is also implicated in the formation of satellite bubbles. The azimuthal shape oscillations described by Schmidt~\textit{et al.}, can be initiated by anisotropic boundary conditions in the liquid as well as with an asymmetric nozzle shape. I measure the $n=3$ oscillatory mode, and observe the nonlinear, highly three-dimensional outcomes of pinch-off with large azimuthal perturbations. These are consistent with prior theory.

\end{abstract}

\pacs{47.55.db, 47.55.df, 02.40.Xx}

\maketitle

\section{Introduction} 

\begin{figure} 
\begin{center}
\includegraphics[width=1.5in]{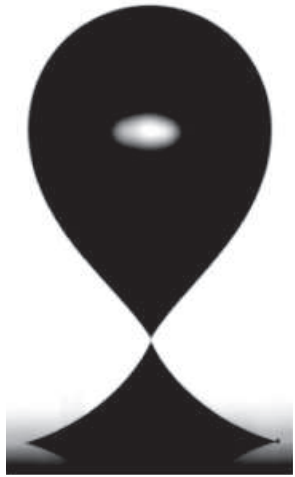}
\end{center}
\caption{Air bubble pinch-off. The bubble is blown from a 6~mm-diameter circular hole cut in a plate that was carefully leveled. The plate reflects the image of the bubble. The bright spot in the dark bubble is due to back-lighting. }
\label{fig:samplepic}
\end{figure}

The breakup of an underwater bubble, whether in a swimming pool, crashing surf,
or even in the processes of a paper mill, exemplifies the rich underpinnings of
even the most fleeting phenomena. In a brief moment, a thick neck of air
collapses to a microscopic radius and breaks --- a process that can be modeled as
the formation of a singularity, where physical quantities, including pressure and velocity, diverge~\cite{eggers94,keller83,shi94,longuet-higgins91}. It was previously
thought that all breakup singularities, from dripping faucets to underwater bubbles,
are universal, so that as the pinching neck of fluid
approaches the singularity, its shape and dynamics are determined solely by the
two fluids' material parameters, and initial conditions are forgotten. Recently, it was discovered that the singularities at the
end of certain pinch-offs instead retain information about their beginnings~\cite{doshi03,keim06}.

In the case of an air bubble pinching off from an underwater nozzle, illustrated by Fig.~\ref{fig:samplepic}, the
singularity has a detailed memory of the bubble shape as pinch-off is
beginning, which may change the character of the singularity formation, or even
disrupt it entirely~\cite{schmidt09,turitsyn09}. These developments raise two central questions: What information is preserved at the singularity? What terms in the dynamics can perturbatively break the symmetry of the singularity and thus disrupt the singularity formation? Studies of such a rich phenomenon as air bubble pinch-off
can thus provide new insights into dynamical singularities, and the
free-surface flows that develop them.

This paper reports on experiments that investigate further what information is preserved at the singularity. It focuses on the way the azimuthal and axial symmetry of pinch-off are broken by initial and boundary conditions~\cite{keim06,schmidt09}, and by the gas inside the bubble~\cite{gordillo07,burton08,gekle10}.

Past and present studies of perturbed breakup begin with models of the fully symmetric pinch-off of a void in an inertial fluid, and the experiments that address these models. Previous investigations of this kind of pinch-off have focused on the shape of the cavity near the
singularity, and the extent to which the evolution of the neck differs from
that of a universal self-similar singularity. These
studies concentrate on determining the radius $\rmin$ of the narrowest part of the neck, measured as a
function of the time $\tau = t_* - t$ remaining until the time of disconnection
$t_*$. The model of Longuet-Higgins~\textit{et al.} uses the approximation that conditions vary slowly with vertical position $z$, so that the pinch-off dynamics can be approximated in a two-dimensional
horizontal slice~\cite{longuet-higgins91}. It shows that the inertia of the water dominates the asymptotic dynamics, so that surface tension,
viscosity, and gas may be neglected, and hydrostatic and Bernoulli pressures
drive the collapse. In this limit, the evolution should approximately follow
the power law $\rmin \propto \tau^{1/2}$, a finding borne out by simulation and
experiment~\cite{longuet-higgins91,oguz93,burton05}. 

Subsequent theoretical and simulational work has shown that when the
three-dimensional shape of the axisymmetric neck is considered, the exponent of
this power law may evolve during the collapse, varying with the logarithm of
$\tau$~\cite{gordillo05,eggers07,gordillo08,gekle09d}. Such changes over many
orders of magnitude cannot be measured in experiment, but these models are
consistent with experiments that have measured a single value for the exponent
that is slightly higher than
0.50, typically 0.56~\cite{bergmann06,keim06,thoroddsen07,bolanos-jimenez08}. The present work finds similar values for the exponent. 

In this paper, I first discuss the methods used to observe and measure pinch-off, to vary the composition of the gas, and to break the azimuthal symmetry of the bubble. This is followed in Sec.~\ref{sec:gasasymm} by results on the role of gas in altering the neck shape and creating satellite bubbles. In Sec.~\ref{sec:vibration}, I report on experiments in which the azimuthal symmetry of the neck is broken by initial or boundary conditions. I observe vibrations of the neck shape, and in some cases, new shapes and topologies.


\section{Experimental Methods} 
\label{sec:methods}

\begin{figure}
\begin{center}
\includegraphics[width=3in]{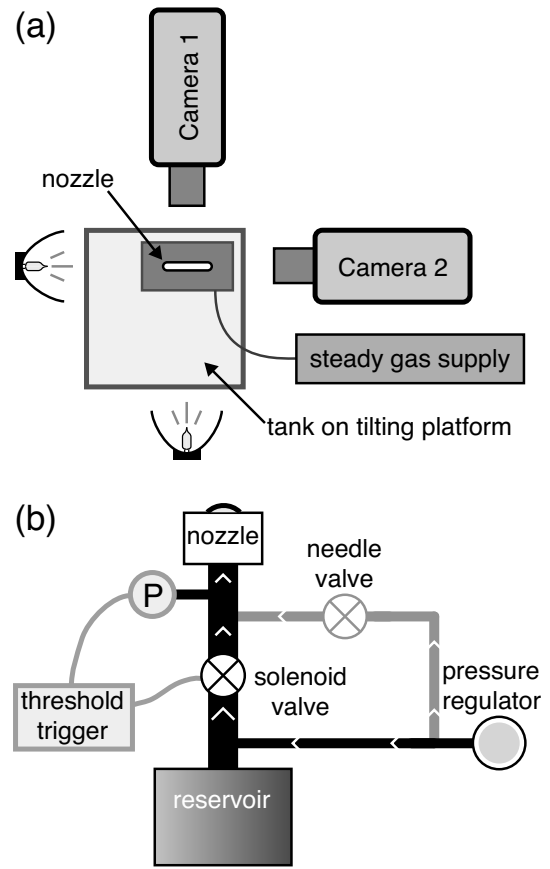}
\end{center}

\caption{\textbf{(a)} Schematic top view of the experimental apparatus. Bubbles
are blown slowly from a nozzle, back-lit, and imaged synchronously with two
high-speed cameras. Several nozzles of different shapes and sizes may be fitted
to the apparatus. \textbf{(b)} Variant of the gas supply of (a) for generating
bursts. Gas is slowly released through a needle valve until the bubble reaches
a preset state; a pressure sensor then triggers the release of the gas held in the reservoir. }

\label{fig:apparatus}
\end{figure} 


My experiment in its simplest form is represented in
Fig.~\ref{fig:apparatus}a. Gas is blown slowly from an underwater
nozzle, adding to the volume of a bubble until surface tension can no longer
hold it at the nozzle, so that it begins to pinch off~\cite{longuet-higgins91}. Gas flows from a pressure-regulated source through a needle valve, and
so is added at a constant rate to the combined volume of the bubble and the tubing connected to the nozzle ($\sim$9~mL). The top of the nozzle is fitted with one of a series of interchangeable aluminum plates,  each with an orifice of a specific size and shape cut out of it, from which the bubble emerges.

Prior experiments have shown that underwater bubble pinch-off has a remarkable sensitivity to any tilt of the nozzle orifice from the horizontal~\cite{keim06}. More generally, the shape of pinch-off is a function of the contact line at the base of the bubble, where the gas-water interface terminates. Therefore the position of this contact line must be controlled carefully. First, the line's position with respect to the nozzle must be fixed. To this end, the inside of the nozzle is machined to reduce its roughness, and made hydrophobic with a coating of
PTFE (``Fluoroglide,'' Saint-Gobain); the PTFE is sanded off the plate's top
surface to make it hydrophilic. This design ensures that the bubble interface stays pinned to the top rim of the nozzle orifice. Second, the orientation of the nozzle itself must be controlled, so that the top of the nozzle is in a plane perpendicular to gravity. A leveling adjustment is therefore performed at the beginning of each experiment, by changing the inclination of the water tank in increments as small as 0.01$^\circ$ until pinch-off has bilateral symmetry when viewed from both the front and side. Because of these measures, azimuthal asymmetry of pinch-off in my experiments is due exclusively to the manufactured shape of the nozzle orifice, or to deliberate tilting.

Bubbles are back-lit and imaged with high-speed video. The use of two cameras,
positioned at a right angle to each other, permits a measurement of the neck's
azimuthal asymmetry. Most experiments reported in this work were imaged with
high-speed video cameras (``Phantom'' v12.1 or v7.3 from Vision Research, Inc.) at up to 183,000 frames
per second with a 1~$\mu$s exposure. A long-distance microscope (``InfiniMax'' by Infinity Photo-Optical) on each camera typically provided a magnification of
3.8~$\mu$m/pixel. Movies were lit with low-infrared continuous
tungsten lights (``DedoCool'' by Dedotec), which prevent significant heating of the water. 
Some single exposures were lit with a 300-ns arc flash (``Palflash'' from Cooke Corp.),
triggered when the disconnecting bubble interrupted a laser beam.

Each resulting movie is analyzed by first selecting a threshold pixel value halfway between the most common black and white pixel values in the movie, then using this threshold to find the positions of the neck edges for each row of pixels, with linear interpolation between adjacent pixels. Pixels where dirt or a camera malfunction would cause erroneous results are excluded from the analysis. This process yields the neck profile, which is the half-width $R$ as a function of height $z$. Random error in the edge position, due to both image noise and the edge-finding process, is manifest as fluctuations in the computed midline of the neck, which would ideally be a straight or gently curving vertical line. These fluctuations were found to be very small, typically well below 0.1 pixels. To find $\rmin$ from $R(z)$, a parabola is fitted to a small region around the minimum. Distance scales were determined by imaging a reference cylinder, 2~mm in diameter, at the focal plane.

To analyze scaling behavior of the minimum neck radius $\rmin$ with time, the $\rmin(t)$ data from each individual movie are fitted with the function:
\begin{equation}
\label{eqn:powerlaw}
\rmin = \beta (t_* - t)^\alpha \mathrm{.}
\end{equation}
To determine the range of exponents $\alpha$ that is consistent with a given movie, a number of trial fits are performed by varying $t_*$ as an independent parameter, and observing the quality of each resulting fit and the value of $\alpha$ that it yields. When judging the quality of a fit, I assume a systematic error for $\rmin$ of $\pm 0.5$~pixels (as opposed to the much smaller random error described above). This reflects the possibility that the edge-tracing algorithm, combined with the optical system, might systematically misjudge the position of the edge. The resulting best value of $\alpha$, and range of $\rmin$ to include in the fitting process, are then used to determine $t_*$ and $\beta$ for each movie in the data set, so that all data may be plotted on a single time axis, $\tau = t_* - t$.


\subsection{Repeatability of quasistatic experiments} 
\label{sec:repeatability}

My normal method of bubble inflation, used in the majority of experiments described here, involves slowly inflating the bubble until its buoyancy overcomes surface tension and it pinches off. When bubbles are inflated in such a quasistatic way, and the position of the nozzle and gas-water interface is well-controlled as described above, pinch-off is initiated with a precise bubble size and shape, so that these quasistatic experiments are remarkably repeatable and reproducible. Nearly all of the plotted data sets in the present work are taken from at least two separate pinch-off events, and the experimental and analysis protocols described here were developed with the goal of ensuring consistency between results obtained from different runs.

Two further sources of non-repeatability that are addressed by my experimental protocols bear mentioning. First, my most sensitive measurements of the asymmetry introduced by a tilted nozzle
suggest that the water currents excited by repeated quasistatic pinch-offs can have a
minute, non-reproducible, systematic effect on this tilting asymmetry at pinch-off. Waiting at least 90~s
between bubbles eliminates this effect. However, these most sensitive experiments on tilting were the only type of measurement where I could find a systematic effect from faster quasistatic bubble production. For all other types of experiments presented in this paper, I was unable to identify any systematic difference between 90~s/bubble and $\sim$10~s/bubble, although a small increase in random variation between bubbles was observed at the faster rate. No data reported in this paper were collected faster than $\sim$10~s/bubble.

Second, external vibrations affect the timing and symmetry of pinch-off, and in some cases, seriously increase the random scatter of resulting data. Experiments were therefore conducted on a heavy optical table supporting as few moving components as possible, and with limited human or mechanical activity in the laboratory. The optical table's isolating suspension system could not be used, however, because it introduced error into the table's orientation with respect to the horizontal.


\subsection{Measuring asymmetry with 2 cameras} 
\label{sec:methods-2cam}

Measuring azimuthal asymmetry with 2 cameras, as in Fig.~\ref{fig:apparatus}, involves finding the difference between the cameras' simultaneous measurements of the neck profile $R(z)$. This subtraction requires transforming the two-dimensional profiles, originally in units of pixels, to a single three-dimensional laboratory coordinate system with units of $\mu$m. Error in each camera's conversion from pixels to $\mu$m (\textit{i.e.}, the magnifications) is the dominant source of systematic error in this transformation. (By contrast, the aforementioned $\pm 0.5$~pixel systematic error in locating the true positions of the neck edges is common to the measurements of both cameras, and thus cancels out in the subtraction.) While the magnification of each camera may be measured by imaging a reference object, this cannot be done with enough precision to then measure an asymmetry much smaller than the neck size. For example, in order to determine that a neck with $\rmin = 100$~$\mu$m is 1~$\mu$m wider in one camera than in the other, each magnification should be known to better than 1 part in 100.

Fortunately, it is possible to measure instead the ratio of the 2 magnifications with sufficient precision. The ratio may be worked out by using movie frames from each camera to measure the separation between the 2 cusps of air (connected to the bubble and to the nozzle) that remain after breakup, as they retract vertically away from the pinch-off point. When repeated over many pinch-off events, this method is equivalent to measuring many reference objects spanning a range of length scales. Furthermore, unlike when using a manually positioned reference object, the calibration and experimental measurements are necessarily in the same focal plane. Each camera's line of sight is kept sufficiently horizontal so that the mismatch between horizontal and vertical magnifications due to projection effects is negligible. This technique also yields the elevation of one camera relative to the other, which is required to compare data from the 2 cameras on a single $z$ axis. Because this calibration process is applied each time the optical system is adjusted, I am able to measure reproducibly asymmetries in the neck shape to better than 1 part in 500, as estimated by comparing similar measurements taken on multiple days.


\subsection{Control of gas} 

Most of the experiments reported here used nitrogen as the gas. To control the gas density and the amount of dissolved gas, some experiments were also performed with He and SF$_6$ gas in a glass vacuum cell that allowed the observation of pinch-off through front and side windows. A vacuum pump with a cold trap was used to lower the pressure to $\sim$10~kPa, though observations were made only at $\sim$20~kPa and above, because of the difficulty of maintaining a stationary contact line at the nozzle when gas density is low (see Sec.~\ref{sec:gasasymm} below). Water was degassed in a separate vessel before the experiment. It was further degassed in the experimental cell by lowering the pressure to $\sim$10~kPa, then returning to atmospheric pressure by slowly bubbling the less-soluble He or SF$_6$ through the nozzle. This procedure was repeated at least once before a final pump-down to $\sim$20~kPa. The nozzle was then leveled according to the procedure described above, after which recorded observations began. Throughout the experiment, pressure was raised solely by introducing gas at the nozzle. This ensured that the quantity of dissolved gas would increase minimally during the experiment, as SF$_6$ and He have low solubility in water (0.007 and 0.009 by volume, respectively, as compared to 0.03 for air~\cite{liquide76}).


\subsection{Bursting apparatus} 
\label{sec:bursting}

To study the pinch-off singularity's memory of perturbations, a nozzle with a non-circular orifice is used to break the azimuthal symmetry of the bubble shape. Most such experiments are carried out with quasistatic inflation to maximize reproducibility (see Refs.~\cite{keim06,schmidt09} and the present work). However, in the course of quasistatic inflation, the smoothing action of surface
tension limits the size of perturbations that one can apply to the bubble
shape. Surface tension must therefore be partially circumvented in experiments that require
larger azimuthal shape perturbations. 

Inflating the bubble in a short time, on
the order of 40~ms, prevents surface tension from smoothing out the
longest-wavelength asymmetries in the bubble shape --- those produced by the
nozzle shape itself. Figure~\ref{fig:apparatus}b shows the apparatus for
producing these rapid ``bursts'' of gas: A reservoir is connected to the nozzle
by a length of plastic tubing, which is constricted by a small solenoid pinch
valve. The pinch valve was chosen to minimize the displacement of gas due to
the valve's operation. The reservoir, whose volume may be adjusted from the
$\sim$1~mL volume of the tubing up to 16~mL, is charged with a pressure
regulator to 0.3--5.1~kPa above the hydrostatic pressure at the nozzle (pressures were set relative to ambient atmospheric pressure; values are quoted in this paper using a typical hydrostatic pressure at the nozzle of 0.4~kPa). In the present work, the reservoir volume is 16~mL unless otherwise specified. Meanwhile, a separate line supplies
gas to the nozzle at a constant, low flow rate, inflating a bubble
quasistatically. In the early stage of this slow inflation, a convex cap
protrudes from the nozzle, corresponding to increasing gas pressure as it
grows. The burst is triggered when the output voltage of a pressure sensor (Honeywell DC002NDC4) reaches a pre-set threshold, corresponding to a particular interface shape. To create the burst, a delay generator momentarily opens the solenoid
valve, triggers the high-speed cameras, and blocks re-triggering for
$\sim$20~s, long enough for the most violent liquid motion from the burst to
subside. The apparatus thus satisfies both requirements for repeatable
bursting: a consistent pulse of gas delivered when the interface is at a
consistent initial shape. In practice, consecutive bursts are very repeatable.
However, perhaps owing to details of pinch valve 
operation, quantitatively reproducing a pinch-off behavior between
runs is difficult, especially for high-pressure bursts, indicating a sensitivity to aspects of the experimental protocol or apparatus that have yet to be fully controlled.



\section{Effects of Gas Density} 
\label{sec:gasasymm}


\begin{figure}
\begin{center}
\includegraphics[width=3.5in]{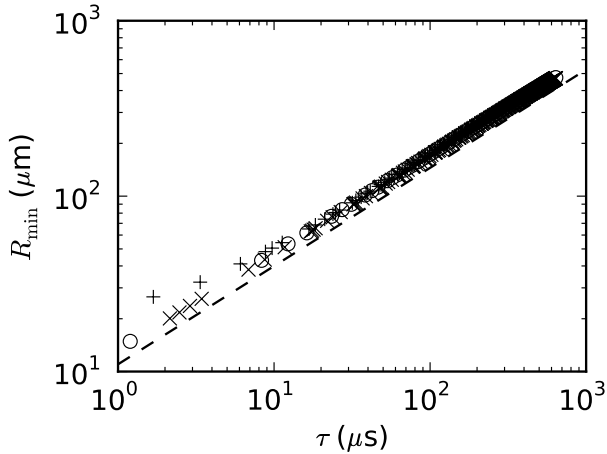}
\end{center}

\caption{Minimum neck radius $\rmin$ as a function of time until break-up $\tau$ for axisymmetric pinch-offs of N$_2$ ($\circ$, density 1.14 g/L, 3 events), low-pressure He ($\times$, 0.0411 g/L, 6 events), and SF$_6$ ($+$, 5.63 g/L, 7 events). $t_*$, and thus $\tau$, is determined for each movie by fitting with Eq.~\ref{eqn:powerlaw} with $\alpha = 0.56$ (dashed line). Only points below $\rmin \sim 120$~$\mu$m were used for fitting. The consistency of these data show that the changing gas density does not strongly change the character of the singular dynamics over most of the collapse --- however, SF$_6$ points below 60~$\mu$m could not be fitted with an exponent of 0.56.}

\label{fig:rmingas}
\end{figure} 

The gas inside the collapsing neck is not required for a model of the singularity at pinch-off~\cite{longuet-higgins91,eggers07,schmidt09}, but prior work has found that it does affect the dynamics nonetheless~\cite{gordillo05,gordillo07,gordillo08,gekle10}. Furthermore, the role of gas in the production of the small satellite bubbles that remain after pinch-off~\cite{gordillo07} is unclear. Here I report on experiments that independently vary the composition of the gas inside the bubble and the ambient pressure of the experiment (thereby varying both gas density and static pressure in the liquid), in order to elucidate the role of gas as it changes the neck shape near the singularity, and leads to the formation of a satellite bubble.

My experiments vary the composition and pressure of the gas, achieving gas densities from 0.039 to 5.5 g/L --- thereby varying the density ratio $\rho_g / \rho_l$ from $3 \times 10^{-5}$ to $4 \times 10^{-3}$, and the Atwood number $(\rho_l - \rho_g) / (\rho_l + \rho_g)$ from 1.000 to 0.989. (All
gas densities are calculated from Ref.~\cite{lide08} for room
temperature.) 

Measurements of the evolution of $\rmin$ for the different gases studied
here confirm that changes to gas composition and density do not strongly
affect the inertial pinch-off dynamics over most of the collapse, so that the effects of gas inertia may be considered as a perturbation to the symmetric void collapse dynamics.
Figure~\ref{fig:rmingas} presents $\rmin$ \textit{vs.}\ $\tau$ data for axisymmetric
pinch-off of 3 gases from a 6~mm-diameter circular nozzle in water: dry N$_2$, He,
and SF$_6$. The initial transient, where surface tension is important, persists down to $\rmin \sim 300$~$\mu$m; below that, pinch-off is an inertial collapse described by Eq.~\ref{eqn:powerlaw}, with $\alpha = 0.56 \pm 0.03$. This exponent was previously found to hold for quasistatic pinch-off from circular nozzles with diameters of 3--8~mm~\cite{keim06}. Notably, the SF$_6$ data in Fig.~\ref{fig:rmingas} differ from the data for other gases, in that points below $\rmin \sim 60$~$\mu$m are described poorly by a power law with $\alpha = 0.56$, suggesting new dynamics at small $\rmin$ due to gas inertia~\cite{gordillo05}. Alternately, the effective power law of the entire
collapse could be higher, in a range centered on $\alpha = 0.58$. The anomaly at small $\tau$ may also be corrected by adjusting the pinch-off time $t_*$, at the expense of introducing a significant deviation near $\tau = 20$~$\mu$s.

\subsection{Effect of gas density on neck shape and vertical motion} 

Rather than being symmetric about its minimum, as one may
expect from a leading-order analysis~\cite{longuet-higgins91}, the collapsing neck of a gas bubble has a slight vertical asymmetry, with a minimum that moves upwards
in time~\cite{bergmann06,bergmann09a,gekle10}. Burton and Taborek recently studied bubble pinch-off for
gas-to-liquid density ratios from $\sim$$10^{-3}$ to 1, and found that this
asymmetry and motion became more pronounced for higher density ratios; they
attributed these effects to a ``jet'' of gas that flows through the neck
minimum~\cite{burton08}. Gekle \textit{et al.}\ simulated and
experimentally measured this same gas flow, and showed that during the collapse of a
void created by the impact of a solid object on water, a slight gas pressure
imbalance between the regions above and below the minimum drives the upward gas
flow~\cite{gekle10}. Late in collapse, the gas reaches velocities much greater than those of the water and exerts a force on the liquid at the narrowest part of the neck,
thus creating the upward motion and asymmetry.

\begin{figure}
\begin{center}
\includegraphics[width=3.3in]{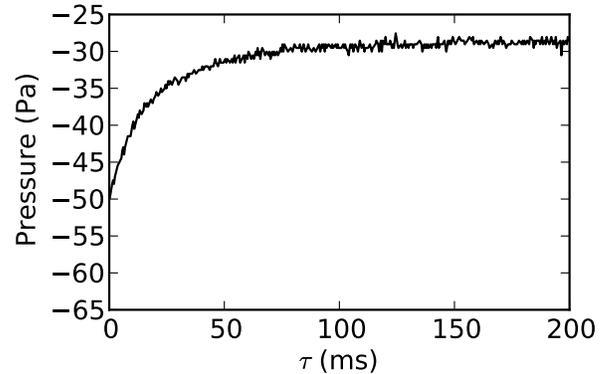}
\end{center}

\caption{
A pressure drop in the upper portion
of the bubble is responsible for the upward gas flow. The pressure in the gas supply line is plotted as a function of the time $\tau$ before pinch-off is seen 
on video, for He at 0.17 g/L as it disconnects from a round, level, 6~mm-diameter nozzle. The pressure drops by 9.0~Pa in the final 10~ms of
pinch-off. Pressure is relative to the value measured 400~ms after pinch-off, once vibrations of the residual interface have died down. The data for SF$_6$ are not significantly different, except for a lag due to the slower speed of sound. 
}

\label{fig:pressuretraces}
\end{figure} 

By placing a gas pressure sensor at the nozzle, I have determined a lower bound on the pressure drop that drives this gas flow through the neck minimum. Gas pressure measured immediately before pinch-off is shown in Fig.~\ref{fig:pressuretraces}. As expected, the pressure gradually decreases as the top
portion of the bubble rises into water with lower hydrostatic pressure; this pressure drop accelerates significantly in the final 20~ms of pinch-off. As in the work of Gekle \textit{et al.}~\cite{gekle10}, the present estimate of the pressure drop responsible for the gas flow is small compared with the absolute pressure in the neck --- 1 part in $10^4$ as measured here, and of the order of 1 part in $10^2$ in the simulations of the prior work~\cite{gekle10}.
My measurement is a lower bound on the magnitude of the final pressure difference: because of the limited time resolution of Fig.~\ref{fig:pressuretraces}, and the location of the pressure sensor at the end of several cm of tubing, some of the rapid change in pressure just before the singularity might be masked. Because this pressure drop is primarily due to the large, rising free surface above the neck minimum, and not to changes near the nozzle, it drives an upward gas flow through the neck minimum. 

Here I show that while this rapid gas flow may play a role in the final moments of pinch-off, it does not explain the vertical motion and asymmetry observed much earlier in the collapse. Instead, it is the gas's role in setting initial conditions, at the onset of the inertial collapse, that allows it to affect vertical asymmetry at these early times. 

\begin{figure}
\begin{center}
\includegraphics[width=3.3in]{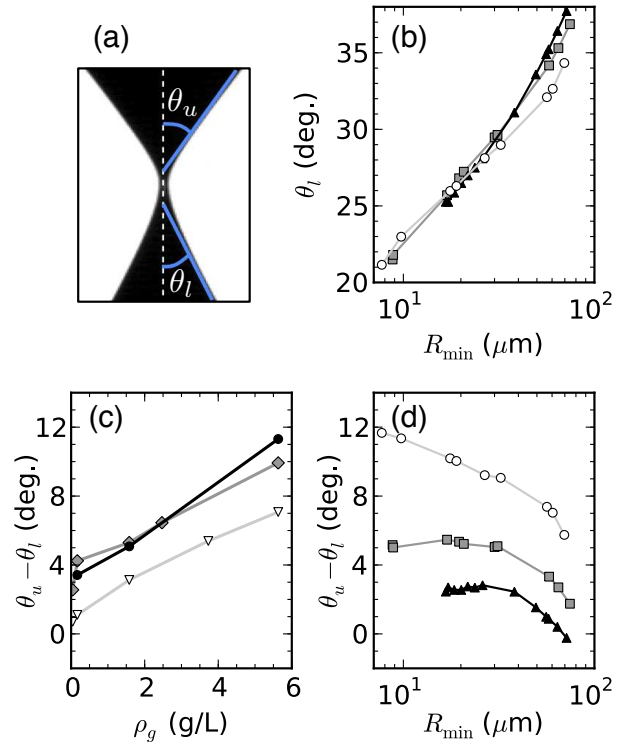}
\end{center}

%

\caption{(Color online) Influence of gas density on the vertical asymmetry of the neck. 
\textbf{(a)} Measurement of cone angles $\theta_u$ and $\theta_l$. For
consistency and repeatability, cone angles are computed by fitting a straight
line to the region of the profile between $7\rmin$ and $12\rmin$ away from the
minimum.
\textbf{(b)} Lower cone angle $\theta_l$ as a function of neck minimum radius $\rmin$, for 3 gas densities: lowest-vacuum SF$_6$ (5.5~g/L, circles), highest-vacuum SF$_6$ (1.5~g/L, squares), and highest-vacuum He (0.039~g/L, triangles). Time proceeds from right to left; thus the neck grows more slender during collapse.
\textbf{(c)} Vertical asymmetry $\theta_u - \theta_l$ as a function of gas density $\rho_g$, sampled at $\rmin = 10$~$\mu$m (circles), 20~$\mu$m (diamonds), and 60~$\mu$m (triangles). The data suggest a small residual asymmetry that is not due to gas inertia.
\textbf{(d)} Vertical asymmetry as a function of neck minimum radius $\rmin$ for the 3 gas densities shown in (b). Higher gas density gives rise to a growth in asymmetry late in collapse. Earlier, asymmetry appears to evolve from an initial condition at large $\rmin$ that is determined by the gas density, but with a trend that is independent of the gas.
}

\label{fig:vertasymm}
\end{figure} 

To investigate the effect of gas density on the vertical asymmetry of the neck shape, I measure and compare the upper and lower cone half-angles of the neck profile, as indicated in Fig.~\ref{fig:vertasymm}a. The angles $\theta_u$ and $\theta_l$ are obtained by fitting a straight line to the regions of the profile for $z$ between 7$\rmin$ and 12$\rmin$ away from the minimum. This measurement technique was chosen for its relative insensitivity to the details of the method, and the repeatability of its results; the measured angles would be smaller near the minimum, owing to $R(z)$ being nearly parabolic in that region. Fig.~\ref{fig:vertasymm}b shows that under all conditions, the profile becomes more slender as $\rmin$ decreases in the approach to pinch-off, consistent with earlier observations~\cite{gordillo05,bergmann06}. 

The asymmetry $\theta_u - \theta_l$, which is apparent in the photograph of Fig.~\ref{fig:vertasymm}a, shows a strong dependence on gas density, as shown in Fig.~\ref{fig:vertasymm}c. Lowering gas density greatly reduces the difference between $\theta_u$ and $\theta_l$, primarily by making the upper shape more slender at pinch-off. Figure~\ref{fig:vertasymm}d shows that gas density makes the greatest difference when the neck is most constricted, at $\rmin \lesssim 20$~$\mu$m and the velocity $\rmindot \gtrsim 7.5$~m/s, as seen in the growth of asymmetry for the highest-density SF$_6$ (5.5~g/L) --- whereas the less dense gases show a neutral or slightly negative trend of asymmetry in this late stage of collapse, consistent with recent simulations of pinch-off with vertical asymmetry and no inner fluid~\cite{herbst11}. The qualitative difference for the highest-density gas at small $\rmin$, as is also seen in the evolution of $\rmin(\tau)$ in Fig.~\ref{fig:rmingas}, suggests that inertial effects change the dynamics late in the collapse. 

\begin{figure}
\begin{center}
\includegraphics[width=3.3in]{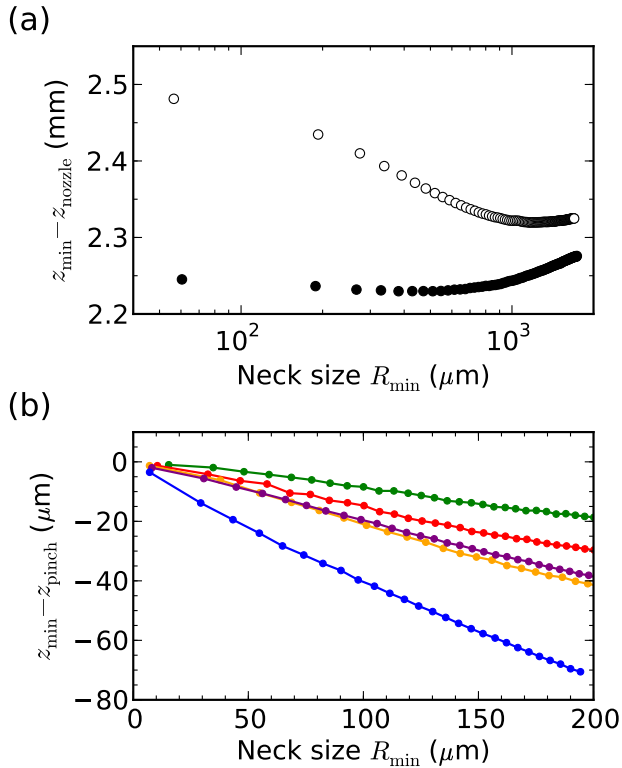}
\end{center}

\caption{(Color online) Upward motion of the neck minimum is driven by a vertical gas flow as a bubble disconnects from a round, level, 6~mm-diameter nozzle.
\textbf{(a)} Height of the neck minimum
above the nozzle as a function of minimum radius $\rmin$ for He at
0.029~g/L ($\bullet$) and SF$_6$ at 5.7~g/L ($\circ$). Increasing gas density causes an upward motion of the neck minimum, which for SF$_6$ is more pronounced and begins at much larger $\rmin$. $\tau$ ranges from 12 to 0.1~ms over the course of this plot.
\textbf{(b)}~Height of the
minimum as a function of $\rmin$ closer to pinch-off, showing a singular
motion at late times that increases with gas density. $z_\textrm{min}$ is compared with its extrapolated value at $\rmin = 0$, $z_\textrm{pinch}$, to make the curves' different slopes more apparent. From top to bottom, the 5 curves
correspond to He at 0.041~g/L, He at 0.16~g/L, SF$_6$ at 1.6~g/L, N$_2$ at
1.1~g/L, and SF$_6$ at 5.6~g/L. The range of $\rmin$ plotted represents the final 100~$\mu$s of pinch-off.}

\label{fig:zmotion}
\end{figure} 

These results suggest that for most of the collapse, vertical asymmetry is due to a
mechanism other than gas flow through the neck minimum, since there is
an effect even at the much lower velocities earlier in the collapse. To explore this
possibility, I turn to a related behavior, the vertical motion of the neck minimum $\zmin$, that is more easily measured at
early times --- including in the transient regime before inertial collapse, when surface tension is still important. Figure~\ref{fig:zmotion}a
shows considerable, gas-dependent motion of $\zmin$ \emph{before} the onset of the inertial collapse at $\tau \sim 1$~ms ($\rmin \sim 600$~$\mu$m). Figure~\ref{fig:zmotion}b shows that this dependence carries over into late times. Around $\rmin \sim 50$~$\mu$m ($\rmindot \sim 3.6$~m/s), the slope steepens further for pinch-off with the highest density, consistent with the results for asymmetry in Fig.~\ref{fig:vertasymm}d. These observations show that much of the vertical asymmetry is present in the initial conditions of the inertial collapse.

I also note that this vertical motion of $\zmin$ causes an increase in the final height of the neck minimum. This effect leads to a problem at low gas densities: as the upward motion is nearly eliminated, a smaller volume of residual gas is left at the nozzle after pinch-off. Because this residual bubble vibrates after it is violently deformed by pinch-off, its flatter shape and lower height mean that the interface may depin and fall below the nozzle rim during the vibrations. Any depinning interferes with the repeatability of pinch-off (see Sec.~\ref{sec:repeatability}), and is the practical limitation on precision low-pressure experiments. For He gas, it causes difficulty at pressures well above the vapor pressure of water, even above 10~kPa.

These results show that gas density has an observable effect on neck shape at all stages of collapse --- even when the minimum neck radius is 2000~$\mu$m, and the maximum liquid velocity is just 0.05~m/s. I argue that this dependence is due to the effect of gas on initial conditions before inertial collapse, including the effect of transient dynamics dominated by surface tension. These initial conditions influence the subsequent evolution, and only at late times is the shape changed by gas flow through the neck minimum. The vertical symmetry of the initial conditions is presumably broken by a combination of buoyancy and geometry (the fixed nozzle below, and only a free surface above) which would be altered by a change in gas density. That this breaks the symmetry of the subsequent collapse may be understood by considering the slenderness approximation discussed above, in which dynamics at different $z$ are roughly independent. In this sense, the inner fluid acts as a perturbation that is remembered by the singular dynamics. By contrast, the effect of gas flow through the neck observed for higher $\rho_g / \rho_l$ and small $\rmin$ requires the introduction of new dynamics late in collapse, and so does not constitute a memory of initial conditions. This latter effect was prominent in the more energetic pinch-offs studied by Gekle \emph{et al.}~\cite{gekle10}, though that work did not investigate the role of the gas density. The work of Burton \emph{et al.}~\cite{burton08}, on a nozzle pinch-off experiment similar to the present one, varies the gas density in the transition from the perturbative regime to pronounced asymmetry and strongly altered dynamics. Those experiments only marginally overlap with the present work, which covers the weakly perturbative regime of extremely small $\rho_g / \rho_l$, and so can describe pinch-off and identify the role of gas in the near-absence of inertial effects.

\subsection{Effect of gas density on satellite bubbles} 

These investigations also give insight into the possible role that gas plays in the
creation of satellite bubbles. These bubbles, of diameter 10~$\mu$m or smaller,
are observed near the point of disconnection after
pinch-off~\cite{burton05,keim06,thoroddsen07}. Their origin is not well-understood. The bubbles may contain vapor or
gas that was released from the surrounding liquid by low dynamic pressure, as
suggested by recent results of Burton \textit{et al.}~\cite{burton08}. Alternately, they
may contain the same gas that comprises the main bubble, due to a mechanism
that would bifurcate the neck minimum just before pinch-off, like that proposed
by Gordillo and Fontelos~\cite{gordillo07}. 
Here gas, which flows rapidly in the vertical direction at the end of collapse, may provide the necessary coupling in $z$ at the small scales of satellite formation. In their model, when gas is forced out of the collapsing neck, there is a stagnation point at the minimum, and gas inertia lowers the dynamic pressure just above and below it, creating the bifurcation. 

Happily, these two possibilities --- satellites from gas in the surrounding liquid, or from the motion of gas already in the neck --- produce two different dependencies on experimental conditions. In the former case, lowering the ambient pressure of the experiment would make the satellites larger, as more gas is drawn from the liquid; degassing the water would make them smaller. In the latter case, lowering the density of the gas would diminish its ability to deform the neck shape, as discussed for the case of vertical symmetry above. Control over the ambient pressure, gas composition, and dissolved gases makes these experiments possible.

In the experiments using degassed water reported here, satellite bubbles were observed at all gas densities. There is a weak dependence of the satellite bubble size on gas density: the bubbles are $\sim$10~$\mu$m diameter for SF$_6$, and at most 3.5~$\mu$m diameter (the size of a pixel) for He. These experiments rule out dissolved gas as a source of satellite bubbles, both because the water was degassed, and because satellite bubbles are not larger at low pressures, when gas would come out of solution more readily. 

A previous effort by Burton \textit{et al.}\ varying gas density instead observed no satellite bubbles at all, suggesting that the use of degassed water prevented satellite formation \cite{burton08}. This is not consistent with the results reported here. Although the mechanism developed by Gordillo and Fontelos~\cite{gordillo07} shows how inertial effects of the gas may lead to the creation of a satellite bubble from gas already present in the neck, this particular mechanism posits a stagnation point at the neck minimum, inconsistent with the strong vertical gas flow discussed above. The results reported here suggest that satellite bubbles are due to the action of the gas already inside the neck, but there is presently no model consistent with these findings.

\subsection{Discussion} 

The above results show that reducing the density of the inner fluid results in pinch-offs with smaller vertical asymmetry. However, over most of the collapse, the vertical asymmetry is due not to rapid gas flow through the neck --- for which the case of zero gas density is easily understood --- but instead to initial conditions, where the effects of gas are weak and depend on the density difference ($\rho_l - \rho_g$). In particular, during quasistatic inflation, when buoyant and capillary forces are balanced, lengths in the system are non-dimensionalized by the capillary length
\begin{equation}
l_c = \sqrt{\frac{\gamma}{(\rho_l - \rho_g)g}},
\end{equation}
where $\gamma$ is surface tension and $g$ is the gravitational acceleration. The analysis of Longuet-Higgins \textit{et al.}~\cite{longuet-higgins91} shows that the bubble shape at the end of inflation has a non-trivial dependence on the dimensionless nozzle size $D / l_c$, however, and so one may expect changes to the conditions at the end of inflation, the transient evolution, and the subsequent inertial collapse, beyond a simple rescaling. 

Fundamentally, the vertical symmetry of pinch-off is broken in two ways: gravity, which causes a hydrostatic pressure gradient, and geometry, specifically the presence of a fixed contact line at the bottom end of the bubble and a free surface at the top. I attempted to change the asymmetry by using an alternate geometry: an apparatus that replaces the free surface at the top of the bubble with a second, downward-pointing nozzle, so that the static bubble neck is created between the two parallel orifices. Pinch-off is initiated by slowly lowering the pressure inside the bubble. The use of this apparatus, however, alters the collapse behavior $\rmin(\tau)$, because surface tension holds the neck open until a much smaller $\rmin$. This leads to poor observations of the singular dynamics. In general, it is more difficult to isolate the contributions of gravity and geometry to symmetry-breaking than it is to identify the role of gas.


\section{Azimuthal memory} 
\label{sec:vibration}


\begin{figure}
\begin{center}
\includegraphics[width=3.0in]{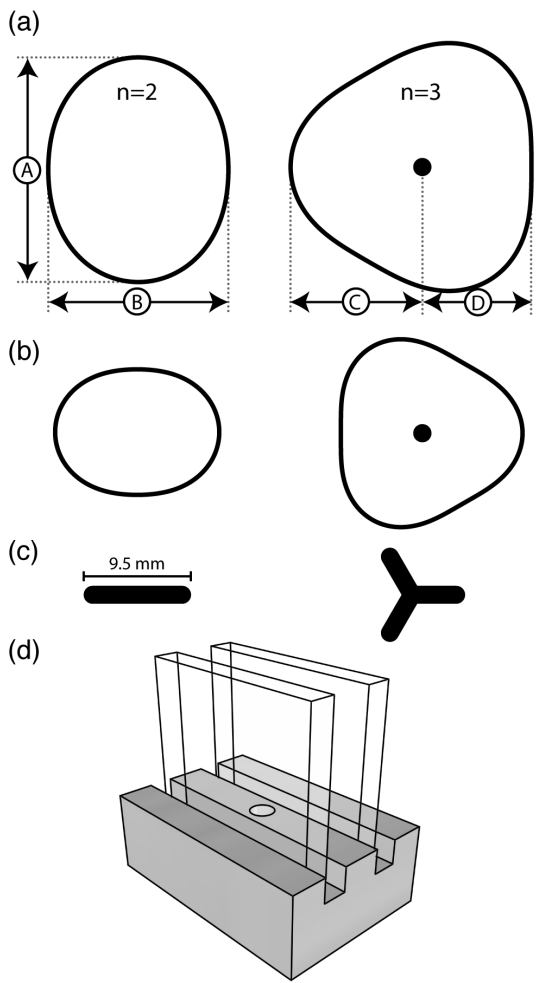}
\end{center}

\caption{\textbf{(a)} Horizontal neck cross-sections showing azimuthal
perturbations~\cite{schmidt09}, and how they are measured. The length $(A-B)/2$,
measured with two cameras, is the $n=2$ asymmetry $\sigtwo$, and the average radius 
$\rbar = (A+B)/4$. The length $(C-D)/2$ is $\sigthree$, where $C$ and $D$ are
measured from the geometric center of the neck, marked by a closed dot; here,
$\rbar = (C+D)/2$. In practice, $C$ and $D$ are determined by assuming that the
geometric center is stationary during the pinch-off, which is valid for the leveled nozzle used here~\cite{keim06}. \textbf{(b)} Cross-sections
at a later time, showing the nature of the shape oscillations (not to scale). The midpoint of the $n=3$ shape, as seen by a camera looking toward the top of the page, is shifted to the right, even though its geometric center has not moved.
The intermediate shape for each mode is a circle. \textbf{(c)} Nozzle orifice shapes used to generate these perturbations. \textbf{(d)} The 2-plate nozzle, used to excite an $n=2$ oscillation by placing two transparent walls near a circular orifice.}

\label{fig:shapes}
\end{figure} 

A key form of memory in bubble pinch-off is the preservation of
azimuthal asymmetry present in the initial or boundary conditions. A perturbation to the initial conditions that breaks
the azimuthal symmetry is remembered in the shape of the neck near
pinch-off~\cite{keim06}. A large class of these perturbations is
addressed by the theory of Schmidt \textit{et al.}~\cite{schmidt09}, which
considers azimuthal perturbations in a single horizontal slice of the
neck, and describes the shape of that slice in terms of harmonic modes on a circle,
\begin{equation}
\label{eqn:shape}
r(\theta, t) = \rbar(t) + \sum_n a_n(t) \cos(\phi_n(t)) \cos(n\theta - \theta^0_n),
\end{equation}
where $\rbar(t)$ is the average radius of the cross-section, $a_n(t)$ is the amplitude of each mode, $\phi_n(t)$ is the phase of the oscillation, and $\theta$ is the azimuthal coordinate. $\theta^0_n$ selects the orientation of the perturbation, and is defined on $[0, 2\pi / n)$.
For $a_n \ll \rbar$, these modes vibrate as $\phi_n(t)$ evolves during collapse, as illustrated for the $n=2$ and $n=3$ harmonic
modes in Fig.~\ref{fig:shapes}. Near the singularity, the phase evolves as
\begin{equation}
\phi_n(t) = \sqrt{n-1} \ln(\rbar(t) / \rbar(t=0)),
\end{equation}
so that as $\rbar \rightarrow 0$, in the limit of small $a_n$, the shape goes through an infinite number of oscillations in finite time. Concurrently, the time derivative ${\dot a}_n \rightarrow 0$, so that any finite perturbation becomes more important as the collapse proceeds --- the absolute amplitude of the perturbation remains constant even as the average bubble radius shrinks to zero.

\subsection{Observations of the $n = 2$ mode} 
\label{sec:n2}

\begin{figure}
\begin{center}
\includegraphics[width=3.5in]{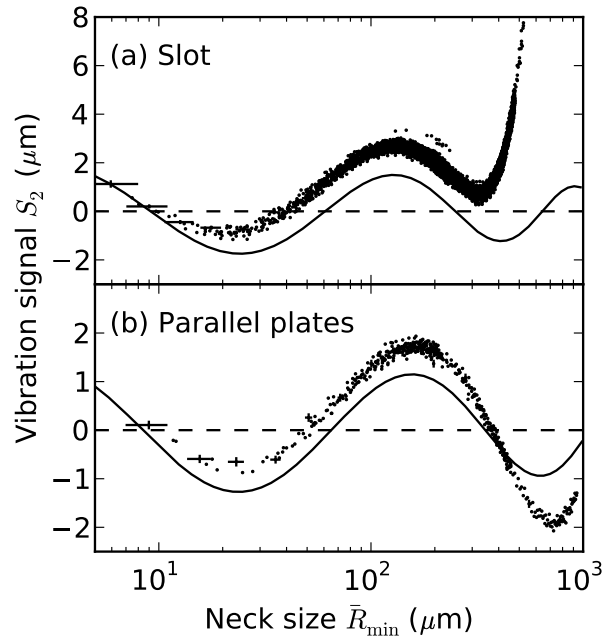}
\end{center}

\caption{Vibrational memory of $n=2$ azimuthal perturbations. The independent
variable $\rbarmin$ is the average radius of the neck at its minimum; time proceeds from right to left. 
Theoretical curves (continuous lines) are
computed according to Schmidt \textit{et al.}~\cite{schmidt09, schmidtthesis08}, taking the
initial condition at $\rbarmin = 128$ and 156~$\mu$m for (a) and (b), respectively.
Representative error bars are shown on a fraction of the points, for clarity.
\textbf{(a)}~The $n=2$ asymmetry $\sigtwo$ oscillates due to
a slot-shaped nozzle (see Fig.~\ref{fig:shapes}). $\sigtwo > 0$ corresponds to elongation perpendicular to that
of the nozzle. 108 separate pinch-off events are plotted.
\textbf{(b)}~$\sigtwo$ oscillates due to parallel vertical plates
on each side of the nozzle. $\sigtwo > 0$ corresponds to elongation parallel
to the walls. 34 events are plotted. The frequency is the same in both experimental configurations, but the phase is slightly shifted and the transient behavior is significantly different.
}

\label{fig:oscdata2}
\end{figure} 

Experiments performed by Schmidt \textit{et al.}~\cite{schmidt09} excited the $n=2$ mode, and showed that the experimentally-observed vibrations agreed with the predictions of this theory. This same agreement is shown for a similar experiment in Fig.~\ref{fig:oscdata2}a. Here, bubbles are blown quasistatically from a slot-shaped nozzle, similar to that shown in Fig.~\ref{fig:shapes}c. (For this particular experiment, I used a shorter 4.6~mm-long version of that nozzle and found that in the quasistatic regime, results are indistinguishable from those of the longer slot.) From the two neck profiles measured by the cameras in each frame of the movie (see experimental methods, Sec.~\ref{sec:methods-2cam}), the average radius $\rbar(z)$ is computed by averaging the two half-widths. The $n=2$ asymmetry $\sigtwo$ is computed at the height $\zmin$ where $\rbar$ is smallest (see Fig.~\ref{fig:shapes}). $\sigtwo$ is plotted here as a function of the minimum radius $\rbarmin = \rbar(\zmin)$, which serves as a clock variable that tracks the time $\tau=t_* - t$ until the singularity. $\rbarmin$ has the advantage that it may be measured independently for each camera frame --- permitting many events to be plotted together without modeling and fitting a $t_*$ for each one. Because $\rbarmin(\tau)$ may be modeled as a power law (Eq.~\ref{eqn:powerlaw}), a plot of these data on a logarithmic $\tau$ axis would have a similar appearance. I have verified that experiments performed with a nozzle rotated by $45^\circ$ show no signal in $\sigtwo$ as expected, since the $n=2$ oscillation is only in orthogonal directions, as in Fig.~\ref{fig:shapes}.

\begin{figure}
\begin{center}
\includegraphics[width=3.5in]{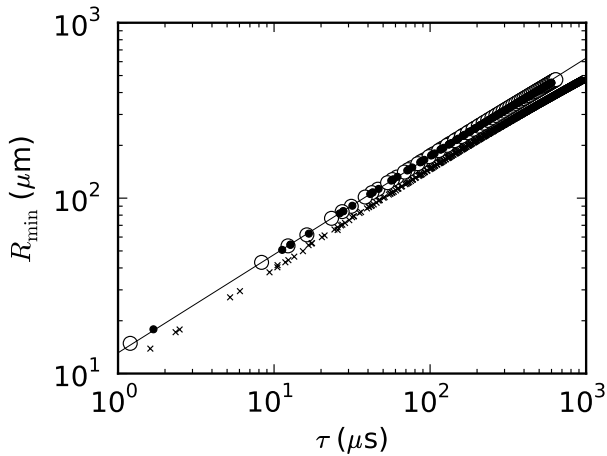}

\end{center}

\caption{Minimum neck radius $\rmin$ as a function of time until break-up $\tau$ for pinch-off from a 6~mm-diameter circular nozzle ($\circ$, 3 events); and minimum average radius $\rbarmin$ of the $n=2$ oscillatory pinch-offs of Figs.~\ref{fig:oscdata2}a ($\times$, 12 events) and b ($\bullet$, 4 events). All plotted events used N$_2$ gas. $\tau$ was determined for each movie by fitting with the power law $\rmin \propto \tau^{0.56}$ (solid line). Only points below $\rmin \sim 120$~$\mu$m were used for fitting.} 

\label{fig:rminmemory}
\end{figure} 

As expected, these azimuthal perturbations do not significantly alter the underlying axisymmetric collapse dynamics. Figure~\ref{fig:rminmemory} shows the evolution of $\rbarmin$ in the $n=2$ oscillatory collapses of Fig.~\ref{fig:oscdata2}, along with the axisymmetric collapse of a nitrogen bubble from a 6~mm-diameter circular nozzle, as was also plotted in Fig.~\ref{fig:rmingas}. The perturbed pinch-off is described well by Eq.~\ref{eqn:powerlaw} with $\alpha = 0.56 \pm 0.03$. (The slot-nozzle data have a smaller prefactor $\beta$ than the axisymmetric case, likely due to the smaller effective nozzle size of the slot-shaped nozzle, whose smallest dimension is only 1.6~mm. The slot-nozzle data are closer to those from a 4~mm-diameter circular nozzle.)

The data in Fig.~\ref{fig:oscdata2}a do not oscillate around $\sigtwo=0$, but instead around $\sim$1.5~$\mu$m. This offset does not appear in the data of Schmidt \emph{et al.}~\cite{schmidt09}. The form of the data in the present work is the result of small refinements in analysis methods, which lead to excellent agreement among several runs of the slot-nozzle experiment, each comprising dozens of movies. The offset persists after exchanging the cameras and their optics, or when using a slot with a perpendicular orientation. This bias is likely due to the sensitivity of the vibrations to asymmetric external flows, and their corresponding pressure fields. Such flows are set up by the early evolution of the asymmetric bubble shape due to surface tension, before the inertial collapse.

Figure~\ref{fig:oscdata2}b directly demonstrates this sensitivity to asymmetric
external pressure fields. I report for the first time that it is possible to excite azimuthal vibrational modes
not only by altering the nozzle shape, but also by imposing asymmetric boundary
conditions in the vicinity of pinch-off. Such boundary conditions cannot affect
the pinch-off dynamics until the water is in motion, after which they create an
anisotropic pressure field by impeding water flow. This excitation is
consistent with the theory of Schmidt \emph{et al.}, which holds that shape
vibrations involve the pressure field of the surrounding water, extending out
to some multiple of the initial neck radius~\cite{schmidt09}. Figure~\ref{fig:oscdata2}b shows
the results of an experiment to demonstrate this principle. Bubbles were released
from a 6~mm-diameter circular nozzle placed between large, parallel vertical
plates spaced 10~mm apart, as pictured in Fig.~\ref{fig:shapes}. The plates are transparent, to permit imaging with
2~cameras at 90$^\circ$ angles as in the slot-shaped nozzle experiment. This
configuration excites the $n=2$ vibrational mode with an amplitude of
$\sim$1~$\mu$m, with the same frequency but a different phase from that in Fig.~\ref{fig:oscdata2}a --- showing that the specific perturbation selects both the phase and amplitude of this linear oscillation. The plates' continual presence as a boundary condition that
alters dynamics, rather than as an initial condition as for the shaped nozzle,
is apparent in the absence of the steep early transient that is seen in the
vibrations from shaped nozzles. Repeating the experiment with the plates removed
restores azimuthal symmetry.

The solid curves in Fig.~\ref{fig:oscdata2} are theoretical $n=2$ curves calculated
with the method outlined in Ref.~\cite{schmidt09}, with the additional refinement of
a viscous term~\cite{schmidtthesis08}. The only adjustable parameters are the empirical amplitude and phase of the oscillation. Each curve describes data only after the neck reaches some
$\rbarmin$~$\sim$~400~$\mu$m, after which the collapse is dominated by inertia,
making the theory applicable. The agreement of the predictions with the data is excellent.

\subsection{Observations of the $n = 3$ mode} 
\label{sec:osc3}

\begin{figure}
\begin{center}
\includegraphics[width=3.5in]{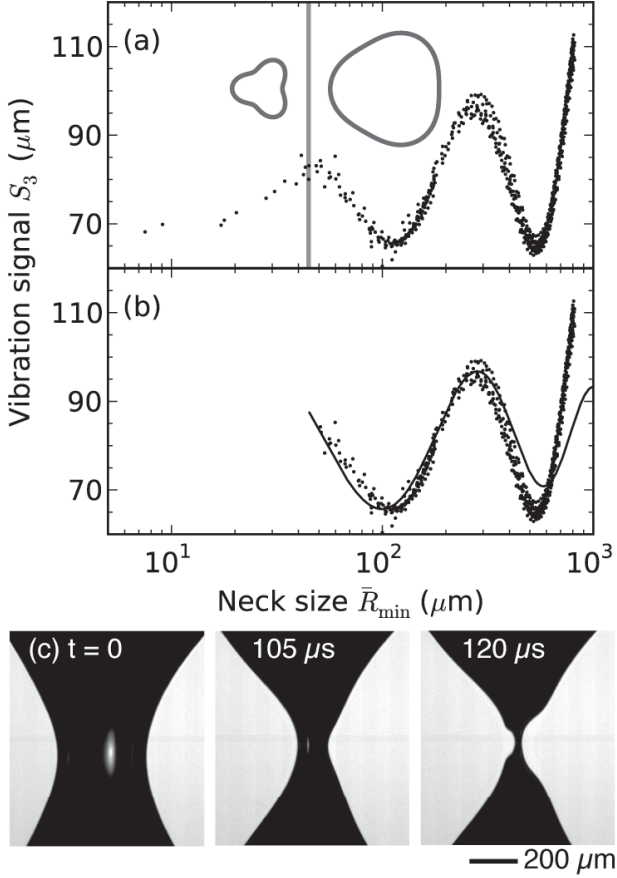}
\end{center}

\caption{Vibrational memory of an $n=3$ azimuthal perturbation due to a 3-armed nozzle (see Fig.~\ref{fig:shapes}).
\textbf{(a)}~Vibration signal $\sigthree$ as a function of neck minimum radius $\rbarmin$. $\sigthree$ is subject to an arbitrary offset, as in Fig.~\ref{fig:shapes}. To better show the consistent phase and amplitude of the
21 events plotted, a different offset is fitted to each event. For $\rbarmin \gtrsim 50$~$\mu$m (shaded vertical line), the neck is sufficiently large compared to the perturbation that the cross-section remains convex (see insets), and so $S_3$ may be accurately measured. To the left of this line, the cross-section is concave and the measurement no longer corresponds to $S_3$. This change in geometry is because the perturbation amplitude remains nearly constant as the neck size decreases.
\textbf{(b)}~The valid portion of the $\sigthree$ data, with a theoretical curve as in Fig.~\ref{fig:oscdata2}. The initial condition was taken at $\rbarmin = 280$~$\mu$m. 
\textbf{(c)}~Frames from one of the movies plotted in this figure. Due to the $n=3$ vibration, the apparent center of the neck oscillates from side to side in time. In the final frame, with $\rbarmin = 20$~$\mu$m, the neck shape above and below the minimum recapitulates this prior side-to-side motion.
}

\label{fig:oscdata3}
\end{figure} 

Here I report the first measurements of the second vibrational mode,
$n=3$~\cite{keim08}. A signal of
this vibration is shown in Fig.~\ref{fig:oscdata3}a; this signal is measured according to
Fig.~\ref{fig:shapes} as an apparent side-to-side motion of the neck. Unlike the $n=2$ data shown in Fig.~\ref{fig:oscdata2}, the $n=3$ observation
comprises more than a full period of the oscillation. The principal development
required for this observation was a means of generating an $n=3$ perturbation
large enough to be detected optically. The smoothing effect of surface tension,
described above in relation to $n=2$ perturbations, is stronger at the shorter
wavelength of $n=3$. Experiments with bubbles inflated quasistatically from an
$n=3$-shaped nozzle (see Fig.~\ref{fig:shapes}) are thus unable
to excite that vibrational mode observably. Instead, I used
small bursts of N$_2$ gas, as described above, to generate sufficiently large
perturbations (in this case, a burst pressure of 7.5~kPa with no reservoir volume). Despite the more rapid bubble inflation, the perturbation is very repeatable; Fig.~\ref{fig:oscdata3} comprises 21 separate events. No $n=2$ signal is observed for an
$n=3$ vibration, and vice-versa, consistent with pure
single-mode vibrations.

Both the analysis of theory and the measurement techniques require $a_n \ll \rbar$. This condition becomes relevant in the $n=3$ oscillation of Fig.~\ref{fig:oscdata3}a: at small $\rbarmin$, the relatively large oscillation amplitude ($a_3 \sim 20$~$\mu$m) 
causes a disagreement between measurement and theory. Crucially, the successful measurement of both $\sigthree$ and $\rbarmin$ requires a cross-section that is entirely convex, as in Fig.~\ref{fig:shapes}. According to Eq.~\ref{eqn:shape}, this condition is satisfied so long as
\begin{equation}
a_3(t) \cos(\phi_3(t)) < (1-\sqrt{3}/2) \rbarmin(t) .
\end{equation}
Using the parameters of the oscillation in Fig.~\ref{fig:oscdata3}a, this predicts a marked transition to concavity at $\rbarmin \sim 50$~$\mu$m, as illustrated by the insets of Fig.~\ref{fig:oscdata3}a. This cutoff is used to replot the data, with the oscillation curve predicted by theory, in Fig.~\ref{fig:oscdata3}b. The theoretical analysis, performed in the same manner as for the $n=2$ oscillation, is in excellent agreement with the valid range of experimental data. The cutoff also means that a measurement of $\rbarmin(t)$ has a small dynamic range and constrains the power law poorly, but it is consistent with Eq.~\ref{eqn:powerlaw} for $\alpha = 0.56$.

Remarkably, the $n=3$ vibration of Fig.~\ref{fig:oscdata3} is also apparent in
still images of the collapsing neck. This is because of the nearly
2-dimensional dynamics responsible for the shape oscillations, which permit the
vibrations to be modeled in a single horizontal slice, independently of those
at other heights~\cite{schmidt09}. Because the initial neck shape perturbation
varies slowly with vertical position $z$, slices of the neck at
different heights are executing nearly the same vibrations but at
different times --- determined by their local average radius $\rbar$. I have found that for all of the regimes of vibrating pinch-off presented in this paper, the oscillation phase and amplitude vary by only $\sim$10$\%$ in the region $|z - \zmin| < 200$~$\mu$m. Thus slices
above or below the neck minimum, having a size $\rbar > \rbarmin$, recapitulate the
oscillations already performed at the minimum. The last frame of
Fig.~\ref{fig:oscdata3}c shows such a neck shape, in which the apparent $n=3$
side-to-side wobbling in time is manifest as a ``kinked'' neck. This kind of neck deformation in discrete steps may account for some of the neck shapes observed with other burst perturbations, as illustrated in Fig.~4a of Ref.~\cite{keim06}. My experiments for $n=3$ have isolated this striking behavior, which should be a generic feature of vibrational memory with odd $n$.

\subsection{Nozzle tilt perturbations} 

\begin{figure}
\begin{center}
\includegraphics[width=3.4in]{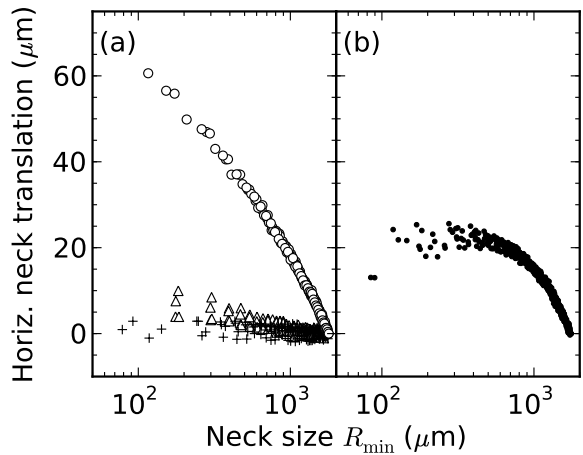}
\end{center}

\caption{
\textbf{(a)} Horizontal translation of the neck minimum as a function of neck size $\rmin$ for a 6~mm-diameter nozzle leveled to within $0.02^\circ$ ($+$), and tilted by 0.1$^\circ$ ($\triangle$) and 0.75$^\circ$ ($\circ$) (4 movies each). Motion is away from the direction of tilt. Unlike that oscillation, translation from tilting is monotonic. \textbf{(b)} By contrast, the signal from placing a single vertical wall 8~mm from the nozzle center ($\bullet$, 14 movies) is weakly non-monotonic, changing at the onset of inertial collapse ($\sim$400~$\mu$m). Here, motion is initially toward the wall.
}

\label{fig:tiltcp}
\end{figure} 

Notably, azimuthal shape vibrations are not excited by the tilting perturbations of quasistatic pinch-off, first
described by Keim \emph{et al.}~\cite{keim06}. Such a perturbation is created
by tilting the nozzle orifice by an arbitrarily small angle from the
horizontal; an effect for a 6~mm circular nozzle can be observed for tilts as
small as 0.05$^\circ$. Figure~\ref{fig:tiltcp}a shows that tilting the nozzle causes the neck minimum to translate 
horizontally away from the direction of tilt during collapse. The final neck shape has a
correspondingly lopsided appearance. At pinch-off, the neck minimum is thinner
in the asymmetric view than in the symmetric one, reminiscent of the
cross-sections produced by an $n=2$ oscillation. But while this perturbation is
an excellent example of the memory of azimuthal asymmetry in pinch-off, I do
not observe any associated oscillatory motion. The tilting memory is also different from the oscillations in that it involves the entire three-dimensional flow around the neck: the neck's horizontal motion has a strong additional dependence on $z$, meaning that the perturbed dynamics cannot be approximated as two-dimensional. By contrast, the shape oscillations near the minimum depend only on the local $\rbar(\tau, z)$, resulting in the behavior described above for $n=3$, wherein the neck above and below the minimum repeats the oscillations recently performed at the minimum.

The tilting perturbation may be compared with the case of placing a single vertical wall $\sim$8~mm from the nozzle center --- a variant of the 2-plate geometry shown in Fig.~\ref{fig:shapes}d.
As shown in Fig.~\ref{fig:tiltcp}b, this perturbation leads to a pinch-off with a weakly non-monotonic side-to-side motion. Unlike with the tilting perturbation, the motion depends primarily on $\rbar(\tau, z)$. In this way, it is like the $n=2$ and $n=3$ oscillations described above. At this time, neither a model of the tilting perturbation nor of the single-wall perturbation has been developed.

\subsection{Vibrations and outcomes of large perturbations} 

\begin{figure}
\begin{center}
\includegraphics[width=3.35in]{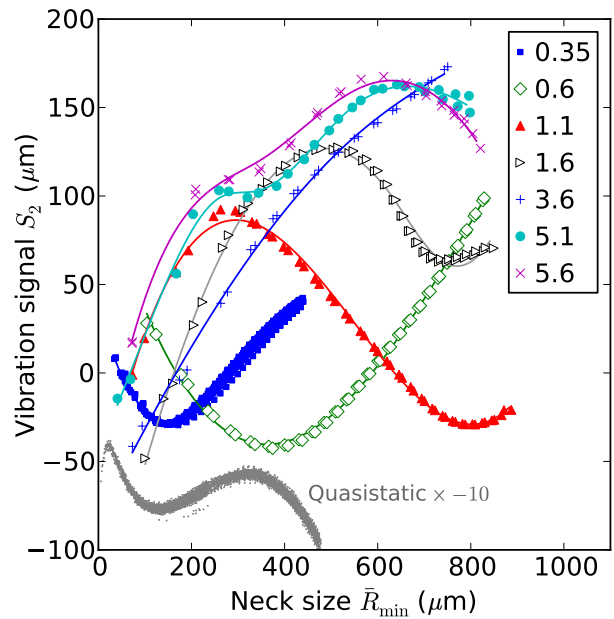}
\end{center}

\caption{(Color online) Varying burst pressure varies the oscillation parameters. Oscillation signal $S_2$ is plotted \textit{vs.}\ average neck radius at the minimum, $\rbarmin$. Each curve represents a burst pressure, labeled in units of kPa. Here $\rbarmin$ has a linear scale, because of the importance of behavior before the inertially-dominated power-law collapse. At lower pressures, each increase in the pressure advances the oscillation phase and increases its amplitude; this effect becomes more complicated, and oscillations less sinusoidal, at higher pressures. As in Fig.~\ref{fig:oscdata2}, $\sigtwo > 0$ corresponds to elongation perpendicular to that
of the nozzle. In the lower left, the oscillation for quasistatic inflation (\textit{i.e.}\ zero burst pressure) from Fig.~\ref{fig:oscdata2}a is plotted with its $S_2$ multiplied by $-10$ and offset for comparison. Increasing the burst pressure gradually changes the phase of the oscillation, and generally increases its amplitude. For each pressure, data points are plotted with a smoothed curve to aid the eye. The points for 0.35~kPa comprise 15 pinch-off events; each other curve comprises 2.
}

\label{fig:giantosc}
\end{figure} 

\begin{figure}
\begin{center}
\includegraphics[width=2.5in]{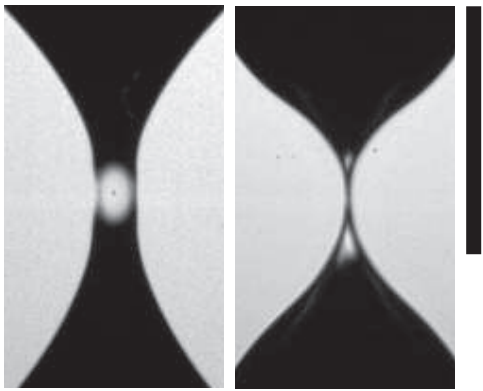}
\end{center}

\caption{Outcome of a small burst of N$_2$ (0.35~kPa) from a slot-shaped nozzle, as seen simultaneously from a
direction perpendicular to the slot (left) and parallel to it (right). In the left image, there is a satellite bubble in the center of the hole. Scale bar is 0.5~mm. }
\label{fig:giantper-splita}
\end{figure} 

\begin{figure*}
\begin{center}
\includegraphics[width=6in]{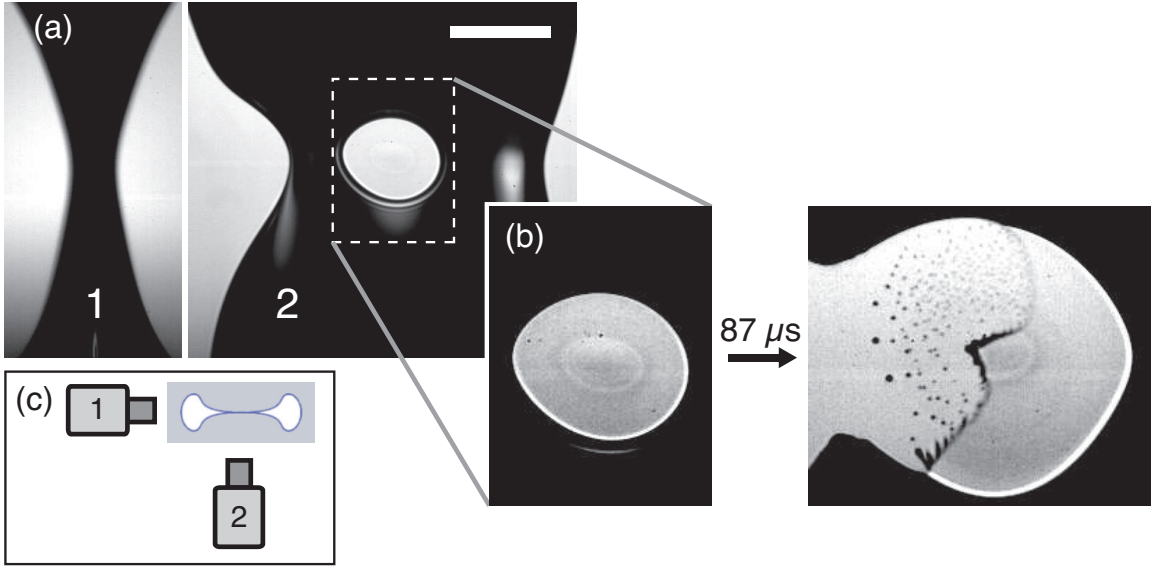}
\end{center}

\caption{\textbf{(a)}~The morphology of Fig.~\ref{fig:giantper-splita} is explained by observing a much stronger burst (5.5~kPa) with the same orthogonal cameras, producing simultaneous images ``1'' and ``2.'' A similar geometry is formed but on a much larger scale, but note that the geometry has changed orientation. Scale bar is 0.5~mm.
\textbf{(b)}~Enlarged and contrast-enhanced detail from (a) shows that the 
``hole'' is actually a $\sim$1~$\mu$m-thick sheet of gas, as indicated by the
presence of Newton's Rings. Right: 87~$\mu$s later, breakup of the thin
sheet proceeds from left to right, leaving hundreds of satellite bubbles.
\textbf{(c)}~Horizontal cross-section of the experiment, showing placement of the cameras used in part (a), and a neck cross-section predicted by
Turitsyn \textit{et al.}~\cite{turitsyn09} that exhibits smooth contact between opposite sides of the neck, consistent with formation of the thin sheet. }

\label{fig:giantper-splitbc}
\end{figure*} 

\begin{figure}
\begin{center}
\includegraphics[width=3.2in]{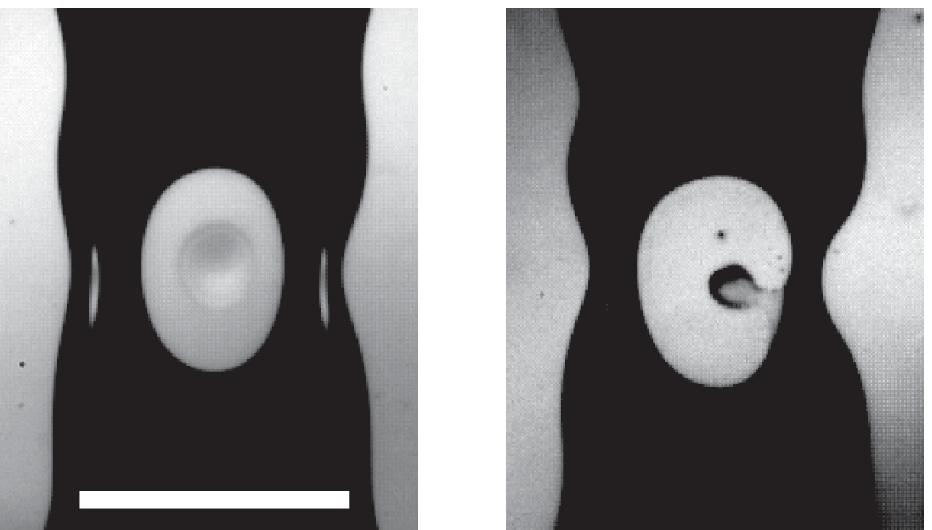}
\end{center}

\caption{300~ns arc-flash exposures taken at different times in the course
of two similar bursts from a slot nozzle. Scale bar is 0.5~mm. Left: Newton's Rings show a thin sheet as in Fig.~\ref{fig:giantper-splitbc}. Right: coalescence proceeds around the perimeter of the sheet, showing a process for producing a single, central satellite bubble as in Fig.~\ref{fig:giantper-splita}, rather than many bubbles, as in Fig.~\ref{fig:giantper-splitbc}. }

\label{fig:giantper-splitd}
\end{figure} 

Schmidt \emph{et al.}\ find that the amplitude of an oscillatory perturbation,
as an absolute length scale, becomes fixed as the overall size of the neck
collapses to zero~\cite{schmidt09}. Thus any finite perturbation will eventually 
dominate the neck shape. This regime has been analyzed by Turitsyn \emph{et
al.}~\cite{turitsyn09}, who show that the cross-sections in this regime are not described by Eq.~\ref{eqn:shape}, which applies only when $\sign \ll \rbar$ as was the case in Fig.~\ref{fig:shapes}. 

The bursting apparatus can create perturbations
large enough to make this behavior visible, and allows access to a variety of oscillation parameters and outcomes by adjusting the pressure of the burst.
Figure~\ref{fig:giantosc} demonstrates this control by showing the $n=2$ oscillations from a series of movies taken with a range of burst pressures. At low pressures, increasing the pressure results in larger amplitudes and advances the oscillation phase. While the data are not recognizable as memory oscillations at higher pressures, the low-pressure bursts still oscillate as though they are in the limit of small perturbations: the data for a burst of 0.35~kPa are remarkably similar to the quasistatic data that are also plotted in Fig.~\ref{fig:giantosc}, even though the perturbation amplitude is larger by an order of magnitude. 

It is only when this 0.35~kPa pinch-off has collapsed down to the scale of the perturbation that there is a dramatic difference, as shown in Fig.~\ref{fig:giantper-splita}. Instead of the pinch-off appearing to end at a single point, two sides of the neck make contact and merge, as in the coalescence of two liquid drops~\cite{keim06}.  The perturbation scale thus cuts off the singularity, replacing simple disconnection at a point with a different topological change~\cite{keim06,schmidt09,turitsyn09}. 
The particular structure and dynamics of the coalescence outcome are detailed in Sec.~\ref{sec:coalescence} below. Large perturbations alter the singularity in every case, though I do not always observe a coalescence --- for some conditions, coalescence may be too small or fast to be imaged, or pinch-off may end with entirely different geometries, such as those suggested by recent simulations~\cite{klfzn-unpub}. The present work, however, focuses on the case of coalescence, because it creates features in camera images (\textit{e.g.}\ the appearance of a hole) that can easily be related to the topology and three-dimensional shape of the neck. 

By varying the oscillation parameters, changing the burst pressure also alters the way in which the singularity
is disrupted. As the neck executes an $n=2$ oscillation, the long axis of a
cross-section changes orientation by 90$^\circ$ for each half-period of the
oscillation. Since increasing the perturbation amplitude will cause it to
disrupt the dynamics sooner in the collapse, one may expect that
amplitude to select the orientation of coalescence events --- an idea more
fully developed by Turitsyn \emph{et al.}~\cite{turitsyn09}. Indeed, in these
experiments, further increasing the strength of the burst results in
coalescences at larger scales, but also changes the orientation of the
coalescence event: while the moderate-sized event of Fig.~\ref{fig:giantper-splita}
takes place along a plane parallel to the slot nozzle, the larger event
of~\ref{fig:giantper-splitbc}a has a similar geometry but in the perpendicular direction. These observations indicate that memory oscillations happen even for large perturbations and $\rbarmin$ that are not strictly within the inertial collapse regime of \textit{e.g.} Fig.~\ref{fig:oscdata2} that Schmidt \textit{et al.}\ analyzed~\cite{schmidt09}. 

\subsection{Coalescence dynamics} 
\label{sec:coalescence}

Examining larger (and therefore slower) coalescence events gives the clearest view of this outcome of strongly perturbed pinch-off. Figure~\ref{fig:giantper-splitbc} indicates that smooth, flat interfaces from opposite sides of the neck are brought into proximity, momentarily forming a thin sheet of gas. Newton's Rings, seen in Fig.~\ref{fig:giantper-splitbc}b, and the estimated volume of the resulting satellite bubbles, indicate that this sheet is on the order of 1~$\mu$m thick. This smooth contact is predicted by the analysis of Turitsyn \emph{et al.}~\cite{turitsyn09} for large $n=2$ distortions, as in the cross-section in Fig.~\ref{fig:giantper-splitbc}c.

Coalescence corresponds to the breakup of the thin sheet of gas. Instead of
observing the process to proceed radially outward from the putative axis of symmetry, as it would for two liquid drops brought into contact~\cite{eggers99,menchaca-rocha01,duchemin03,case09}, one can see one of two
related behaviors: For the larger perturbations, as in
Fig.~\ref{fig:giantper-splitbc}, a front of disintegration sweeps across the thin
sheet, leaving hundreds of satellite bubbles. This behavior is reminiscent of the analysis of inviscid two-drop coalescence by Duchemin \emph{et al.}, who find a rapid sequence of breakups as the air sheet retracts~\cite{duchemin03}. For smaller perturbations, however, as in
Fig.~\ref{fig:giantper-splitd}, the sheet simply retracts around its perimeter,
leaving a central piece of the sheet that typically condenses into a single
satellite bubble. Coalescence proceeds rapidly, and with the exception of large
events as in Figs.~\ref{fig:giantper-splitbc}, is not observed in more than one
frame of video, suggesting a timescale of approximately 5~$\mu$s or less for
events such as Figs.~\ref{fig:giantper-splita} and~\ref{fig:giantper-splitd}. The implied rewetting velocity at which the front of breakups moves across the sheet is several m/s, consistent with balancing liquid inertia with the action of surface tension at the edge of the 1~$\mu$m-thick sheet~\cite{duchemin03}. Finally, an instance of another remarkable feature of large perturbations is the waviness of the outside neck edges in Fig.~\ref{fig:giantper-splitd}. This shows the neck shape's recapitulation of the large-amplitude $n=2$ oscillations at the minimum, analogous to Fig.~\ref{fig:oscdata3}c. These coalescence outcomes illustrate the rich three-dimensional structures that may result from simple perturbations at the beginning of pinch-off.


\section{Conclusions} 

In this paper, I have presented experiments that further reveal some of the possibilities for perturbing the fully non-universal fluid pinch-off of gas bubbles underwater. Shape vibrations may be excited not only by changing the nozzle shape, but by breaking the azimuthal symmetry of nearby boundary conditions. The model of Schmidt \textit{et al.}~\cite{schmidt09} applies beyond the $n=2$ mode previously studied, agreeing with my observations of the $n=3$ vibrations that it predicted. When the perturbation scale is increased to a size that can be imaged more easily, these same dynamics are seen to produce fully 3-dimensional shapes, which require an alternative imaging technique such as X-rays~\cite{turitsyn09,klfzn-unpub}. These outcomes --- including the remarkable coalescence dynamics that can leave hundreds of satellite bubbles --- show not only the ability of this type of implosion singularity to retain information, but its ability to create complex shapes over a range of length scales from a simple perturbation on a single length scale.

The gas within the bubble is likewise a remembered perturbation, in that it contributes, by way of initial conditions and transient dynamics, to the \emph{vertical} asymmetry and position of pinch-off. With sufficient gas density, the effects of rapid gas flow through the neck minimum are also observed as new dynamics late in the collapse. My experiments further implicate inertial gas flow in the creation of satellite bubbles, but since the gas flow rules out the stagnation point that is required by the theory of Gordillo and Fontelos~\cite{gordillo07}, the underlying mechanism by which gas bifurcates the neck minimum remains an open question. An understanding of satellite bubble formation might be of particular importance to applications in which one size of bubble is needed exclusively.

The results presented here are representative of the richness of bubble pinch-off that has been revealed through the recent work of many investigators. This complexity underscores the potential for further insights into this remarkable class of singularities that retain information.

\begin{acknowledgments} 
I thank Sidney Nagel and Wendy Zhang for abundant insight and support. 
I also thank Justin Burton, Daniel Herbst, Heinrich Jaeger, Lipeng Lai, Robert Rosner, Laura Schmidt, and Tom Witten for many helpful discussions. This work was supported by NSF grant No.\ DMR-0652269. Use of facilities of the University of Chicago NSF-MRSEC and the Keck Initiative for Ultrafast Imaging are gratefully acknowledged.
\end{acknowledgments} 

\bibliography{references}

\end{document}